\documentclass[final,5p,times,twocolumn]{elsarticle}

\usepackage{lineno}
\usepackage{hyperref}
\usepackage{tabularx} 
\usepackage{multirow}
\usepackage{amsmath,amsfonts}
\usepackage{xcolor}

\journal{--}

\begin{document}

\begin{frontmatter}

\title{Predicting Rate of Cognitive Decline at Baseline \\Using a Deep Neural Network with Multidata Analysis\\{\small\color{red} This study has been accepted and published in the SPIE Journal of Medical Imaging. https://doi.org/10.1117/1.JMI.7.4.044501}}


\author{Sema Candemir\corref{mycorrespondingauthor}, Xuan V. Nguyen, Luciano M. Prevedello, Matthew T. Bigelow, Richard D.White, Barbaros S. Erdal, \\for the Alzheimer’s Disease Neuroimaging Initiative$^1$}
\address{Laboratory for Augmented Intelligence in Imaging of the Department of Radiology, The Ohio State University College of Medicine}
 \fntext[myfootnote]{Data used in preparation of this article were obtained from the Alzheimer’s Disease Neuroimaging Initiative (ADNI) database (adni.loni.usc.edu). As such, the investigators within the ADNI contributed to the design and implementation of ADNI and/or provided data but did not participate in analysis or writing of this report. A complete listing of ADNI investigators can be found at:
\url{http://adni.loni.usc.edu/wp-content/uploads/how_to_apply/ADNI_Acknowledgement_List.pdf}}




\begin{abstract} 

\vspace{0.2cm}
\hspace{-0.35cm}\textit{Purpose:} This study investigates whether a machine-learning-based system can predict the rate of cognitive decline in mildly cognitively impaired patients by processing only the clinical and imaging data collected at the initial visit. 

\vspace{0.2cm}
\hspace{-0.35cm}\textit{Approach:} We built a predictive model based on a supervised hybrid neural network utilizing a 3-Dimensional Convolutional Neural Network to perform volume analysis of Magnetic Resonance Imaging and integration of non-imaging clinical data at the fully connected layer of the architecture. The experiments are conducted on the Alzheimer’s Disease Neuroimaging Initiative dataset. 

\vspace{0.2cm}
\hspace{-0.35cm}\textit{Results:} {Experimental results confirm that there is a correlation between cognitive decline and the data obtained at the first visit. The system achieved an area under the receiver operator curve (AUC) of $0.70$ for cognitive decline class prediction.}

\vspace{0.2cm}
\hspace{-0.35cm}\textit{Conclusion:} To our knowledge, this is the first study that predicts “slowly deteriorating/stable” or “rapidly deteriorating” classes by processing routinely collected baseline clinical and demographic data (Baseline MRI, Baseline MMSE, Scalar Volumetric data, Age, Gender, Education, Ethnicity, and Race). The training data is built based on MMSE-rate values. Unlike the studies in the literature that focus on predicting Mild Cognitive Impairment-to-Alzheimer`s disease conversion and disease classification, we approach the problem as an early prediction of cognitive decline rate in MCI patients. 
\end{abstract}


\end{frontmatter}

\section{Introduction} 
Mild Cognitive Impairment (MCI) is an intermediate stage between Cognitively Normal (CN) and Alzheimer’s Disease (AD)~\cite{markesbery2010neuropathologic}. The patients in the MCI phase have a varied prognosis such that the cognitive functions of some MCI patients deteriorate, while others remain stable or improve~\cite{petersen1999mild}\cite{qarni2019multifactor}. Although there has not been any successful treatment to reverse cognitive decline, to date, therapy to decelerate its progression is likely to be most beneficial if it is applied early~\cite{lee2019predicting}\cite{cummings2007disease}. In this study, we investigate whether a machine learning-based system can predict the “rate of cognitive decline” in patients with diagnosed MCI by processing only the clinical and imaging data obtained at the initial visit.

Prior studies have reported on biomarkers and the prediction of MCI-to-AD conversion~\cite{suk2013deep}\cite{eskildsen2015structural}\cite{moradi2015machine}\cite{lee2019predicting}\cite{tabuas2016prognosis}. Unlike earlier studies, we investigate the feasibility of predicting the “rate of cognitive decline” in MCI patients at the first visit by processing only the baseline MRI and routinely collected clinical data. We built a deep-learning-based predictive model that integrates imaging and non-imaging demographic and clinical data in the same neural network architecture. The system consists of 3 main inputs: 1) MRI brain images, 2) scalar volumetric features, and 3) demographic and clinical data. MRI brain scans are provided to the network as sequential DICOM images and processed through a 3-Dimensional Convolutional Neural Network (3D-CNN). The scalar volumetric features extracted using FreeSurfer~\cite{FreeSurfer} represent selected brain substructures and included total intracranial volume, whole-brain volume, and regional volumes of the hippocampus, entorhinal cortex, fusiform gyrus, and medial temporal lobe; this scalar data is integrated into the system at the fully connected layer of the architecture. The demographic and clinical information included in the neural network architecture are the ones that are routinely collected at the initial clinical visit, and include age, gender, years of education, ethnicity, race, and baseline mini-mental score examination score.  The proposed predictive model is illustrated in Figure~\ref{Fig:SystemPipeline}. 

We supervised the predictive model with the change in “Mini-Mental State Examination (MMSE) scores~\cite{arevalo2015mini}\cite{harrell2000severe}” with the MCI subjects grouped clinically according to (i) slow cognitive decline, and (ii) fast cognitive decline. The resulting model processes the clinical data obtained at the baseline visit and predicts the patient’s cognitive condition as either “slowly deteriorating/stable” or “rapidly deteriorating”. The analysis is performed on publicly available Alzheimer’s Disease Neuroimaging Initiative (ADNI) dataset (c.f., Section~\ref{Data}).


\section{Materials and Methods}~\label{Method}  \subsection{Data}~\label{Data} The data used in this study were obtained from the ADNI~\cite{ADNI}, which is an ongoing multi-center study. The primary goal of ADNI has been to test whether serial MRI, positron emission tomography (PET), other biological markers, and clinical and neuropsychological assessment can be combined to measure the progression of MCI and early AD. The subjects in the dataset were diagnosed as AD, MCI, Subjective Memory Concern, or Cognitively Normal (CN) based on Mini-Mental State Examination (MMSE) scores. The enrolled subjects received multiple longitudinal follow-up visits over several years, as specified by the ADNI protocol. For our research, we utilize data on ADNI patients who were clinically diagnosed as MCI at their baseline visits. A total of 569 subjects were included. The demographics and clinical characteristics of the subjects are summarized in Table~\ref{Table:Demographics}.

\begin{table}[ht]
\footnotesize
\begin{center}
\begin{tabular}{ c | c }
 Label & MCI \\
\hline
\hline
Number of Patients                  &  569 \\
Age - mean [range]                   &  76 [55 - 92] \\
Gender [Female - Male]           &  F:241  M:328 \\
Education - mean [range]        & 16 [6 - 20]\\
Ethnicity                                       & Not Hispanic / Hispanic  \\
Race                                               & White - Black - Asian \\
Baseline MMSE score - mean [range]    & 28 [23 - 30] \\
 \end{tabular}
 \end{center}
 \caption{Characteristics of the study subjects. MMSE- Mini-Mental State Examination. MCI - Mild Cognitive Impairment.}
 \label{Table:Demographics}
 \end{table}

The MMSE, which is a 30-point test, is a cognitive assessment tool~\cite{arevalo2015mini}\cite{harrell2000severe}, and we used the rate of decline in MMSE scores to supervise the system. Changes in MMSE scores in follow-up visits demonstrate the patient’s condition in terms of cognitive capabilities. A decrease in MMSE score reflects deterioration in cognitive capabilities; if a patient’s cognitive capability is stable, the MMSE scores remain relatively stable.

\begin{figure}[h!]
	\centering
		\includegraphics[height=7cm]{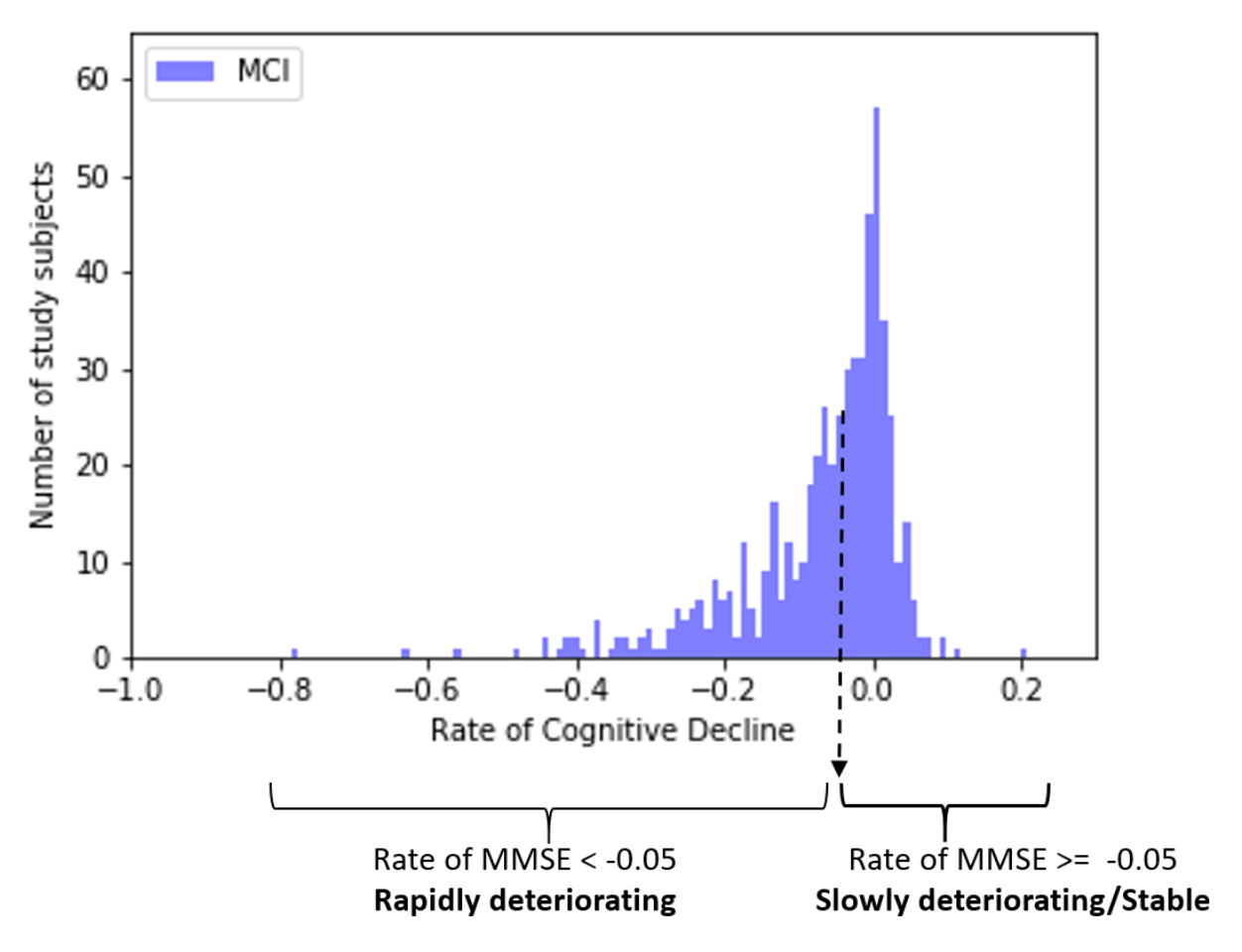}
		\caption{ The rate of cognitive decline distribution of the study subjects}
		\label{Fig:MMSE_Dist}
\end{figure}

We model the change in MMSE scores by fitting a line to the scores obtained at follow-up visits. The slope of the line indicates the rate of cognitive loss. A patient who has faster cognitive deterioration would have a higher absolute value of slope. A slope close to zero indicates that the cognitive decline is stable. In this document, the “Rate of Cognitive Decline” term will refer to the slope of the decline. The predictive model is binary. Therefore, the rate of cognitive decline is converted to binary variables using a threshold of -0.05 points/month, such that progressive rapidly deteriorating level of cognition is defined as a rate of decrease exceeding 0.6 points/year.  This threshold approximates the mean and the median rate change for this cohort. The rate of cognitive decline distribution of the study subjects is shown in Figure~\ref{Fig:MMSE_Dist}.

\subsection{The System Pipeline}
\begin{figure*}[ht]
	\centering
		\includegraphics[height=9cm]{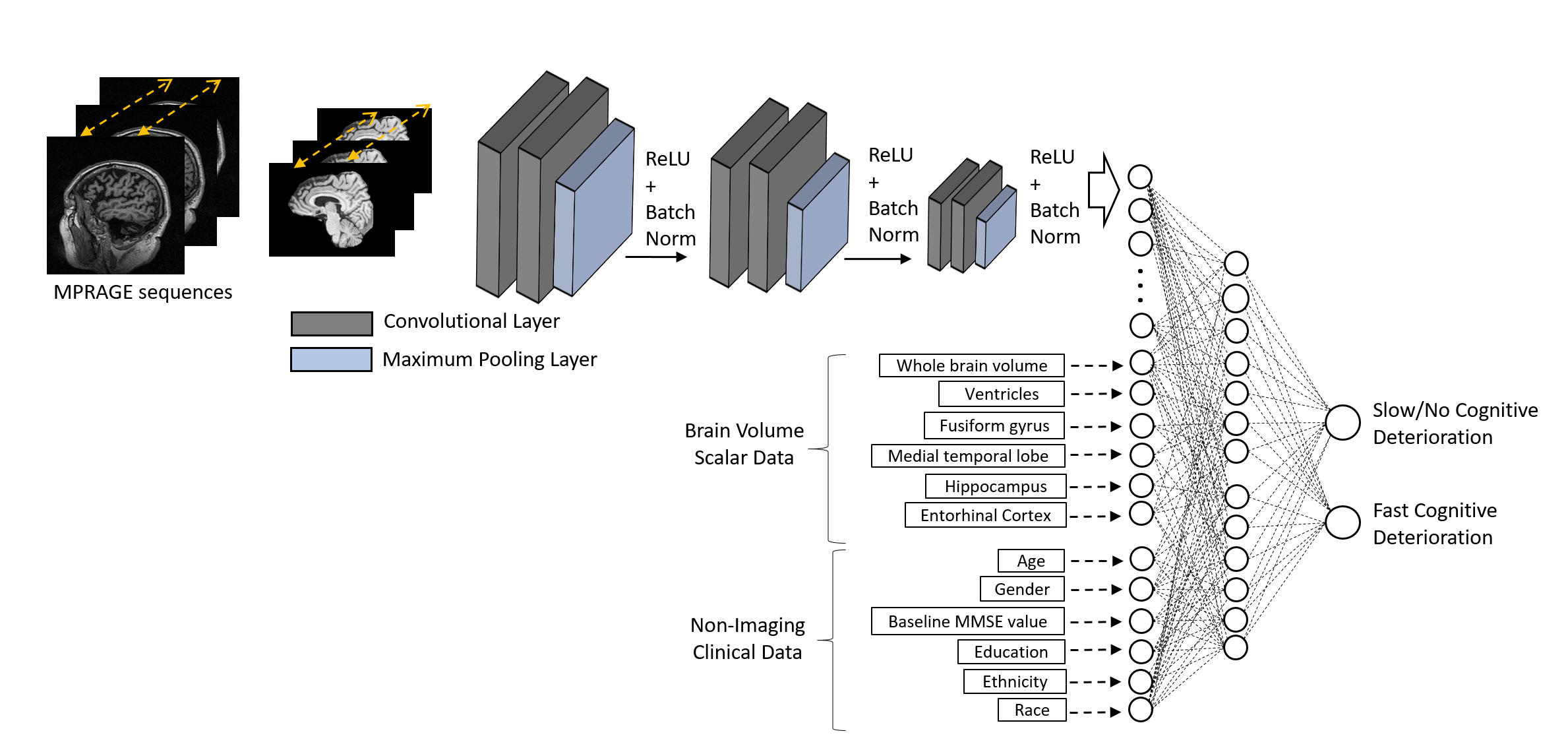}
		\caption{An illustration of the hybrid prediction system. MPRAGE: Magnetization Prepared Rapid Gradient Echo. MRI: Magnetic Resonance Imaging. ReLU: Rectified Linear Unit. Batch Norm: Batch Normalization.}
		\label{Fig:SystemPipeline}
\end{figure*}

The predictive model learns the mapping function from input data to the target output. Let $V$ be the imaging sequence, $D$ be the corresponding clinical data, $y$ be the target class, and $f(.)$ represent the mapping function between input data and output labels. The model can be formulated as
\begin{equation}
  y_i = f(V_i, D_i)
\end{equation}
for each subject i in N, where N is the number of patients with MCI in the training data. Clinical data include age, gender, baseline MMSE score, education, ethnicity, and race. We also use brain volumes as supporting scalar features, which are computed with an open-source library (FreeSurfer) that analyses and visualizes structural and functional neuroimaging data~\cite{FreeSurfer}. Specifically used scalar features are whole-brain volume and regional volumes of the hippocampus, entorhinal cortex, fusiform gyrus, and medial temporal lobe. The brain volumes of each subject are available in the ADNI~\cite{ADNI}. We pose the problem as a supervised classification task, with training subjects classified into two groups based on the rate of MMSE  (c.f., Figure~\ref{Fig:MMSE_Dist}). The output variable  $y \in (0,1)$ denotes the target classes, 0 represents “slowly deteriorating/stable” class, and 1 represents “rapidly deteriorating” class. The proposed system is illustrated in Figure~\ref{Fig:SystemPipeline}.

\subsection{Pre-processing} We apply pre-processing techniques to each MRI volume $V$ and corresponding clinical data $D$ before the training. The MRI sequences are skull-stripped, which includes removal of non-cerebral tissue (calvaria, scalp, and dura)~\cite{esmaeilzadeh2018end}. The skull-strip algorithm, which is based on U-Net architecture~\cite{ronneberger2015u} trained on skull-stripping datasets~\cite{Deepbrain}, reduces the processing size of volumes, thereby increasing computational speed during the training. After the skull-strip, we have applied MRI scale standardization~\cite{nyul2000new} to mitigate the intensity differences between the MRI sequences. 

The neural network architectures require the inputs to be scaled in a consistent way for a stable and faster convergence. Therefore, we normalize the images, scalar volumetric features, and demographic and clinical data into the range between 0 and 1. The scalar regional volume features are divided by each subject’s whole-brain volume size for normalization. The demographic and clinical data contains categorical values (e.g., gender and ethnicity) that are converted into numeric data. The range for the numeric demographic data is between 0 and 1 to ensure numerical stability. 

\subsection{Model Configuration of the Neural Network} The deep learning algorithm is based on a supervised neural network that has a hybrid architecture with two main components: (i) a 3D Convolutional Neural Network (3D-CNN) that learns the brain morphology and patterns, and (ii) integration of scalar volumetric features and non-imaging data (demographic and clinical information) at the fully connected layer.

\subsubsection{Convolutional Neural Network} The 3D-CNN processes MRI scans models the patterns and structures in brain volume. Earlier layers of the model capture the low-level features of brain details, while higher-level layers learn abstract features. The layout of the 3D-CNN architecture is employed from~\cite{esmaeilzadeh2018end} that is proposed for MRI analysis for AD/CN classification. The architecture consists of three batches of convolutional layers with kernels of $3\times3\times3$ elements. Each batch contains two convolutional layers with 32, 32, 64, 64 and 128, 128 filters, respectively. The batches are then followed by a batch normalization that mitigates the over-fitting and improves the system generalization by normalizing the output of the convolutional layer~\cite{ioffe2015batch}. After batch normalization, max-pooling layers with $2^3$, $3^3$, and $4^3$ sizes are used for feature reduction and spatial invariance. The architecture uses ReLU (Rectified Linear Unit) activation that introduces nonlinearity to the system~\cite{lecun2015deep}. The output of the deepest convolutional layer is flattened and fed to the fully connected layer. The architecture parameters are listed in Table~\ref{Table:ArchitectureParams}. 

\subsubsection{Integration of Scalar Volumetric Features and Non-Imaging Demographic and Clinical Data with CNN} The clinical and demographic information presumably contains additional information that would help the classification decision. To incorporate the non-imaging data for assessment of its impact, we have changed the standard CNN architecture. The convolutional part of CNN is the feature extraction component of the architecture; the fully connected layer part is the classifier component. The output of the final pooling layer, which holds the imaging features, is flattened and fed into the fully connected layer. The flattened imaging features and non-imaging features create a concatenated vector as an input to the dense layer. The rest part of the architecture is the classifier component of the hybrid system, trained with this vector to form the final prediction model. The concatenated dense layer is then followed by a dropout layer~\cite{srivastava2014dropout} in which the system temporarily ignores randomly selected neurons during the training to prevent the system from memorizing the training data with the intent to decrease overfitting. The final layer is another dense layer with a softmax activation function with two nodes that provide probabilities for “slowly deteriorating/stable” class and “rapidly deteriorating” class. The architecture parameters are listed in Table~\ref{Table:ArchitectureParams}. 

\begin{table*}[tbp]
\footnotesize
\begin{center}
\begin{tabular}{ c | c  |c | c | c || c}
\bf Layer(Type)    &\bf Input Size (output shape) &\bf \# of Filters& \bf Level of Pooling& \bf \# of Parameters \\
\hline
\hline
 \bf Input (MRI)&  $130\times116\times83$ & & &0  \\
 & & & &  \\
 Conv3D  &  $130\times116\times83$ & 32& --& 896 \\
Conv3D  &  $130\times116\times83$ & 32& --& 27680 \\
 Batch Norm &  $130\times116\times83$ & 32& -- & 128  \\
 Max Pooling 3D &  $65\times58\times41$ &32& 2&0  \\
 & & & & & \\
 Conv3D  &  $65\times58\times41$ & 64& -- &55360\\
 Conv3D  &  $65\times58\times41$ & 64& -- &110656\\
 Batch Norm &  $65\times58\times41$  &64& -- &256  \\
 Max Pooling 3D &  $21\times19\times13$ &64&3& 0 \\
  & & & & & \\
 Conv3D  &  $21\times19\times13$ & 128& -- &221312  \\
Conv3D  &  $21\times19\times13$ & 128& -- &442496  \\
 Batch Norm &  $21\times19\times13$  &128& -- & 512 \\
 Max Pooling 3D &  $5\times4\times3$ &128&4&0 \\
  & & & &  \\
Flatten & 7680 & & & 0 & \bf Input (Metadata)\\
Dense & 512 & & & 3932672 & 12\\
\hline
\hline
\bf Concatenate & 524  & & & 0 & \\
Dropout & & & &  & \\
Dense & 256 & & & 134400 & \\
Dropout & & & &  & \\
Dense (Out) &2 & & & 514 & \\
 \end{tabular}
 \end{center}
 \caption{The architecture parameters of the proposed model. Each row represents a layer, and the input of a particular layer is the output of the previous layer. There are two input layers: (1) processing MRI sequences, (2) processing meta-data. The meta-data input layer is added to the architecture at the Dense layer through a Concatenate function.}
 \label{Table:ArchitectureParams}
 \end{table*}

\subsection{Addressing Overfitting} The voxel-based CNNs are prone to over-fitting due to high-dimensional data, large number of parameters, and relatively small number of cases to optimally train the system~\cite{litjens2017survey, esmaeilzadeh2018end, arbabshirani2017single}. To address the relatively low number of patients, we utilized augmentation strategies. We flipped MRI volumes such that left and right hemispheres are reversed~\cite{esmaeilzadeh2018end} and randomly tilted at less than $5^{\circ}$. We have also employed the regularization techniques of dropout~\cite{srivastava2014dropout} and weight decays~\cite{krogh1992simple} in order to increase the generalization capacity of the model. The parameters of dropout and weight decays are listed in able~\ref{Table:Parameters}.

\section{Experiments}~\label{Experiments}
\subsection{Implementation Details} The dataset used in the study consists of 569 subjects with MPRAGE (MRI) scans and corresponding clinical data  (c.f., Section~\ref{Data}). We perform $5$-fold cross-validation to reduce the performance difference due to relatively small size datasets and provide more robust generalization performance. At each fold, $60\%$ of the dataset is used to train the model, $20\%$ is used for model validation, and $20\%$ of the dataset is used to test the model.

\begin{table}[tbp]
\footnotesize
\begin{center}
\begin{tabular}{ c | c }
   Parameter              & Value \\
\hline
\hline
Processing dimension of each MRI volume & $116 \times 130 \times 83$ voxels\\
Optimizer  &  Adam~\cite{kingma2014adam} \\
Learning Rate  & $0.00005$ \\
$\beta_1$ & 0.9\\
$\beta_2$ & 0.999\\ 
$\epsilon$ & $10^{-8}$\\
Loss Function & categorical cross-entropy\\
Batch Size & 16\\
Drop-out keep rate & 0.5\\
$L2$ Weight Regularizer ~\cite{krogh1992simple}  Kernel coefficient & 0.5 \\
$L2$ Weight Regularizer  Bias coefficient & 1\\
Early stopping max epoch & 400\\
Early stopping patience epoch & 20 \\
 \end{tabular}
 \end{center}
 \caption{Implementation Details - Parameters }
 \label{Table:Parameters}
 \end{table}

The training parameters are listed in Table~\ref{Table:Parameters}. We train the model using Adam optimizer~\cite{kingma2014adam}, which provides faster convergence due to the velocity and acceleration components.  As a training strategy, we monitor the model performance and use two early stopping callbacks to stop the training before the model begins to overfit~\cite{goodfellow2016deep}. We set a large epoch value (c.f., Table~\ref{Table:Parameters}, max epoch 400) as an upper bound iteration. The number of training iteration is decided automatically based on the model performance on the validation and training set. If the validation loss has started to increase during the training process, the system triggers the early stopping callback. If the validation loss continues to increase for another 20 iterations, then the system stops the training. The continuous increase in validation loss is an indication of overfitting. The second callback is monitoring the training accuracy. If the training accuracy reaches the maximum value, the early stopping callback stops the training due to an indication of no further improvement in the model. The weights are randomly initialized from scratch. 

The model is developed in Python (version 3.6.8) using Tensorflow Keras API (version $2.1.6-tf$) and trained on an Nvidia Quadro GV100 system with 32GB graphics cards with CUDA/CuDNN v9 dependencies for GPU acceleration.


\subsection{Evaluation} We built 3 models: (i) an imaging model based on a 3D-CNN that processes brain MRI, (ii) a hybrid model that combines the 3D-CNN component with brain-volume scalar data and demographic and clinical information, and (iii) a model that processes brain-volume scalar data and demographic and clinical information. We assess the models' prediction performance in terms of accurately classifying the cognitive decline on a test dataset at each test fold and average the evaluation metric scores across all the models. The performance metrics used in the study are Sensitivity, Specificity, Accuracy, PPV, NPV, and AUC. Table~\ref{Table:scores} lists the performance metrics.


\subsubsection{Imaging module prediction performance} 
\begin{figure*}
\centering
\includegraphics[width = 8cm]{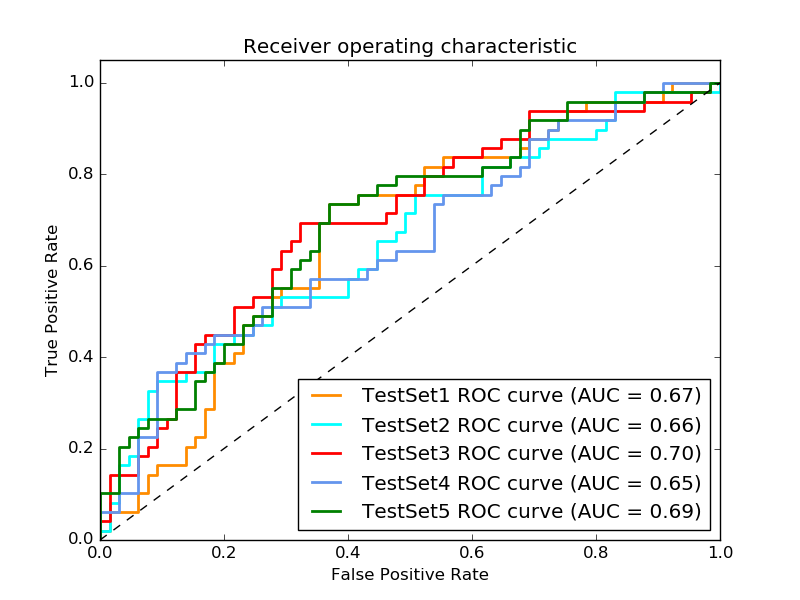}
\includegraphics[width = 8cm]{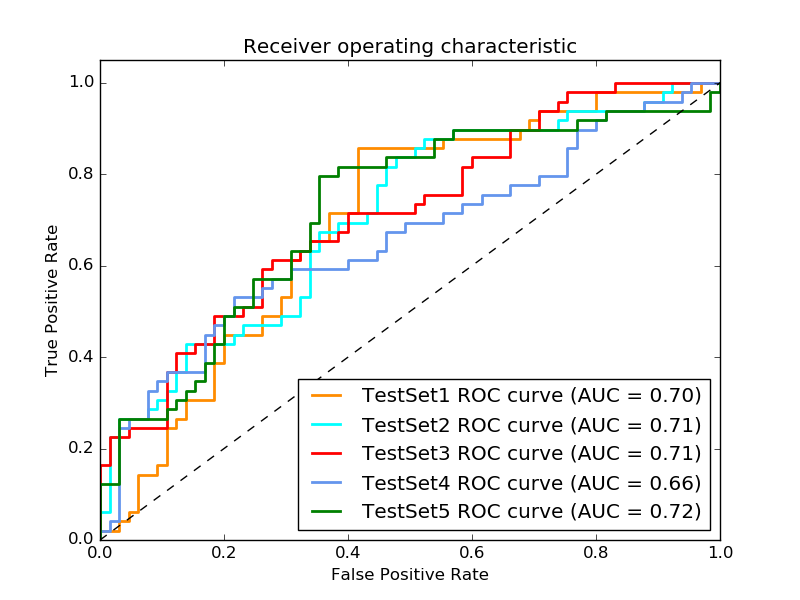}	
	
(a) \hspace{7cm}(b)

\caption{The plots depict the system performance for predicting cognitive-decline class. (a) The predictive model processed only MRI sequences with 3D-CNN; average AUC is 0.67. (b) The hybrid predictive model is based on MRI sequences with 3D-CNN, brain-volume scalar data and non-imaging clinical data; the average AUC = 0.70.} 
		\label{Fig:ROC}
\end{figure*}

The correlation between the morphological changes in the brain (e.g., parenchymal volume loss) and AD is known~\cite{van2006hippocampal}\cite{karas2004global}. Based on a prior study~\cite{lin2018convolutional}, (i) MCI subjects have medium atrophy of hippocampus; (ii) the brain morphology in non-converters is similar to brain morphology in CN, and converters are more similar to AD, and (iii) converters have more severe deterioration of neuropathology than non-converters. Due to the correlation between the pathological changes in brain morphology and the AD stages, we first measured how much we could predict the pace of the cognitive decline of patients by processing only the baseline MRI scans through a 3D-CNN. The system achieved $0.67$ AUC for predicting the cognitive-decline class by processing only baseline MRI sequences. The Receiver Operator Characteristic (ROC) curve for this experiment is shown in Figure~\ref{Fig:ROC}.(a).

\subsubsection{Hybrid model prediction performance} The hybrid model processes the MRI sequences, brain volume scalar data, and demographic information (age, gender, years of education, ethnicity, and race). Table 4 lists the performance scores obtained with the proposed system in terms of mean and standard deviation across the cross-validated folds. The system achieved an accuracy of $63.3\%$, with a PPV of $56.9\%$, sensitivity of $60.8\%$, specificity of $65.2\%$, and NPV of $69\%$ at threshold 0.5. The average AUC is $0.67$. Adding the brain volume and demographic information as scalar values to the system increased the system performance from $0.67$ AUC to $0.70$ AUC as shown in Figure~\ref{Fig:ROC}.(b).

\begin{table*}
\scriptsize
\begin{center}
\begin{tabular}{ c | c|c | c | c | c | c |c  }
Metric th $=$ 0.2  &  Metric th $=$0.3  &  Metric th  $=$ 0.4  &  Metric th $=$ 0.5  &  Metric th  $=$ 0.6  &  Metric th $=$ 0.7  &  Metric th  $=$ 0.8  &  Metric th $=$ 0.9\\
\hline
\hline
 0.630 $\pm$ 0.024   & 0.644 $\pm$	0.026 & 0.646$\pm$ 0.008      & 0.633 $\pm$ 0.024   & 0.630	$\pm$ 0.034   &  0.628 $\pm$ 0.047     &  0.595 $\pm$	0.047    &   0.549	$\pm$ 0.052  \\
 0.619 $\pm$ 0.045  & 0.611	$\pm$ 0.046  &  0.597	$\pm$ 0.022   &   0.569 $\pm$ 0.026  & 0.560	$\pm$ 0.035   &  0.550 $\pm$ 0.042      & 0.520	$\pm$ 0.039  &   0.488	$\pm$ 0.036  \\
 0.380 $\pm$	0.103  & 0.490 $\pm$	0.069  & 0.551 $\pm$	0.072   &   0.608	$\pm$ 0.077 & 0.657 $\pm$	0.068  &  0.755 $\pm$	0.072     & 0.820 $\pm$	0.088      &   0.890 $\pm$	0.065    \\
 0.818	$\pm$ 0.076  & 0.760 $\pm$ 0.071   &  0.717	$\pm$ 0.059   &   0.652	$\pm$ 0.053   & 0.609 $\pm$	0.061  &  0.532	$\pm$ 0.073    & 0.425 $\pm$ 0.106     &   0.292	$\pm$ 0.095 \\
 0.638	$\pm$ 0.023  & 0.665	$\pm$ 0.019   & 0.681 $\pm$	0.018   &   0.690	$\pm$ 0.032  & 0.703	$\pm$	0.037  & 0.744	$\pm$ 0.060      & 0.766	$\pm$ 0.068      &  0.787 $\pm$	0.099 \\
\end{tabular}
\end{center}
 \caption{The hybrid model prediction performance. Training: $60\%$, Validation: $20\%$, Test: $20\%$ of ADNI baseline set. (FN=False Negatives, FP=False Positives, NPV=Negative Predictive Value,  PPV=Positive Predictive Value, TN=True Negatives, TP=True)}
 \label{Table:scores}
 \end{table*}

\subsubsection{Brain Volume Scalar Data and Non-Imaging Clinical Data Prediction Performance} 
\begin{figure*}
\centering
\includegraphics[width =8cm]{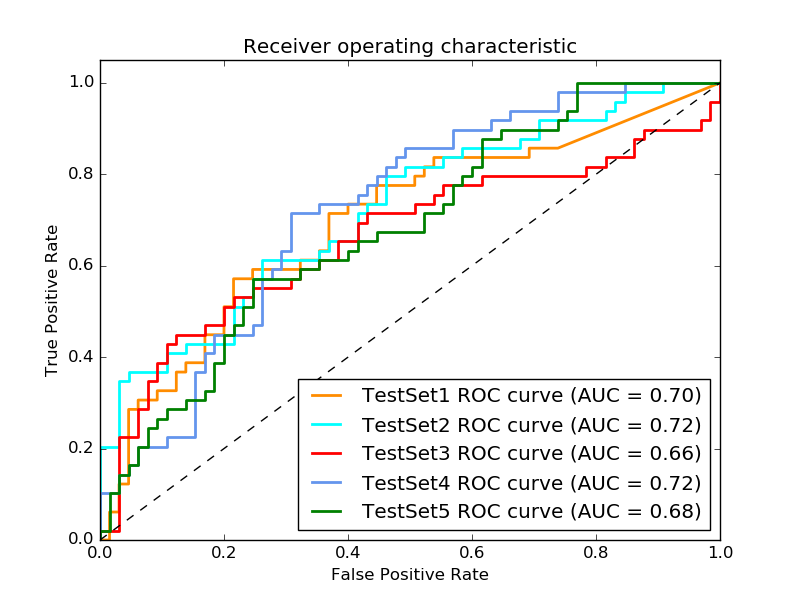}
\includegraphics[width = 8cm, height = 6.2cm]{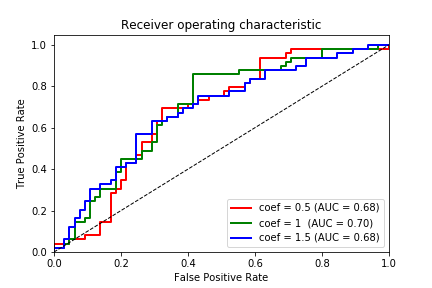}

(a) \hspace{7cm}(b)

\caption{(a) The predictive model processes only scalar data (brain-volume and non-imaging clinical data); average AUC is 0.70. (ROC: Receiver Operator Characteristic) (AUC: Area Under the Curve). (b) An illustration of the hybrid prediction system. MPRAGE: Magnetization Prepared Rapid Gradient Echo. MRI: Magnetic Resonance Imaging. ReLU: Rectified Linear Unit. Batch Norm: Batch Normalization. } 
		\label{Fig:meta}
\end{figure*}

The voxel-based convolutional neural networks are prone to over-fitting due to high dimensional data, large number of parameters, but relatively low number of subject to optimally train the system~\cite{litjens2017survey, esmaeilzadeh2018end, arbabshirani2017single}. Although we utilize several regularization techniques, we still observed over-fitting due to the 3D-CNN module of the hybrid system. In this experiment, we remove the 3D-CNN module of the hybrid model and run the experiments only using brain$-$volume scalar data with non-imaging clinical data. The system achieved $0.70$ average AUC for cognitive decline class prediction as shown in Figure~\ref{Fig:meta}.(a).

\subsection{Integration of Components}
The literature has several techniques to combine data from different resources. We summarized these studies in Section~\ref{sec:Comparison}. We utilized CNN-based architecture as a classifier and combined the imaging features with non-imaging data at a dense layer in a straightforward way, as in \cite{esmaeilzadeh2018end}\cite{dimitriadis2018random}. Note that imaging features or non-imaging features should not dominate the training. To our knowledge, there is not any CNN-based study that adjusts the effects of modules on the prediction results. External weights can be used to adjust the contribution of one module over the other modules. However, these weights are additional hyper-parameters of the system and can be decided on train/validation subsets. In order to observe how different weights affect the final decision, we have conducted an additional experiment. We have multiplied scalar input values (scalar brain volumes and demographic data) with external weight (coefficient value) and kept the MRI imaging weights the same. The external weight adjusts the contribution of the nodes to the classifier decision. The coefficient set is {0.5, 1, and 1.5}. Figure~\ref{Fig:meta}.(b) shows the performance of the model with different coefficient values. 

\subsection{Comparison with the Literature}\label{sec:Comparison}
\textit{AD-MCI-CN Classification.} The correlation between the morphological changes in the brain (e.g., parenchymal volume loss) and AD has been known for years \cite{van2006hippocampal}\cite{karas2004global}. The literature has several studies with quantitative analysis of brain MRI to assess AD (e.g., classification of AD vs. CN) \cite{long2017prediction} \cite{esmaeilzadeh2018end}. These studies measured the volumes, cortical thickness, or shape of various structures such as the hippocampus \cite{lin2018convolutional} or the whole brain \cite{esmaeilzadeh2018end} to assess the disease and the severity of the disease as a percentage of volume. For example, in \cite{esmaeilzadeh2018end}, a 3D-CNN-based framework is proposed to learn the imaging characteristics of AD and CN through convolutional layers. The model is further modified to diagnose MCI, the prodromal stage of AD. In our study, instead of anatomy-disease correlation, we focus on anatomy-function correlation. A comprehensive review of AD detection/classification can be found in~\cite{vieira2017using}.

\textit{MCI-to-AD Conversion.} Many researchers have attempted to predict the conversion of MCI to AD using the correlation between the morphological changes in the brain and disease progression~\cite{lin2018convolutional}. The volumetric analysis of the brain (especially the hippocampus and entorhinal cortex) produces satisfactory results in predicting conversion to AD~\cite{frisoni2010clinical}~\cite{mcevoy2011mild}. One of the most commonly employed classifiers is the Support Vector Machine (SVM)~\cite{moradi2015machine}~\cite{rathore2017review}. Recent studies utilize deep-learning-based approaches using neural network classifiers~\cite{lin2018convolutional}\cite{spasov2019parameter}; they show that predicting progressive MCI or detecting MCI patients who later progress to AD is still a goal of ongoing research~\cite{lee2019predicting}~\cite{tabuas2016prognosis}.

\textit{Hybrid Models.} The clinical and demographic information contains additional data that contributes to the algorithm decision. To our knowledge, the algorithms that incorporate clinical data into MRI data results are limited~\cite{qiu2018fusion}. In \cite{esmaeilzadeh2018end}, age and gender information are concatenated with imaging features in a 3D-CNN architecture through additional nodes at the fully connected layer. In~\cite{lin2018convolutional}, a CNN was trained with local patches extracted from the hippocampus and combined with FreeSurfer brain data. The algorithm extracted imaging features through a CNN architecture and processed imaging features and FreeSurfer brain data using principal component analysis following by the Lasso regression algorithm. The processed features were provided as input to a NN algorithm that combined these features. In \cite{qiu2018fusion}, the authors proposed a multi-modal fusion model to classify MCI and CN cases. The study employed two Multi-Layer Perceptron (MLP) architectures to train non-imaging data and 2D-CNN to train the imaging data. The predictive model processed the test scores of mini-mental state examinations, the Wechsler memory scale for logical memory, and MRI sequences. The predictions from each NN block were then combined using majority voting. Another interesting study that combined baseline MRI with baseline cognitive test scores was proposed in~\cite{moradi2015machine}. The cognitive scores used in the study were Rey's Auditory Verbal Learning Test, Alzheimer's Disease Assessment Scale cognitive subtest, MiniMental State Examination, Clinical Dementia Rating Sum of Boxes, and Functional Activities Questionnaire. The MRI, age, and cognitive measurements were integrated as input features to a random forest classifier. Another hybrid method was proposed in~\cite{lu2018multimodal} that combined MRI and FDG-PET images at multiple scales within a NN framework. Six independent deep NNs processed different scales of image sequences. Another NN fused the features extracted from these first 6 DNN. The algorithm was proposed to classify AD and NC cases. One of the most recent studies is \cite{spasov2019parameter} that combined MRI sequences, demographic, neuropsychological, and APOe4 genetic data to predict MCI patients who have a likelihood of developing AD within 3 years. The study combined the imaging data with non-imaging data at the fully connected layer of their proposed deep-learning-based architecture. We list the recent studies that combined different sources of information in Table~\ref{Table:Literature}. Several other detailed comparison tables can be found in~\cite{vieira2017using}\cite{spasov2019parameter} and in \cite{moradi2015machine}.

\begin{table*}[h!]
\footnotesize
\begin{center}
\begin{tabular}{ c | c | l| c | c | c }
 Study            &  Method                         &   Data  Source        & Brain Region &  Objective  &Performance \\
\hline
\hline
  This Study  &   3D-CNN                         & Baseline MRI +                        & Whole Brain &  Predicting            & AUC: $0.70$     \\
	                       &  Multi-modal                  &  Baseline MMSE +            &                         &   Fast Decliners & Acc: $63.3\%$\\
                       &                                            &  Demographic data  +           &                         &                            & Sens:$60.8\%$\\
									    &                                            &   Baseline Scalar Volume                                                       &                         &                            & Spec:$65.2\%$  \\
\hline
 Lee \textit{et al. }\cite{lee2019predicting} &  Recurrent NN            &  Baseline MRI +                                    & Regional  &  MCI-to-AD                                &  AUC: $0.86$\\
                                                                               &   Multi-modal             &  Demographic data +                              &    Hippocampus        & Conversion               & Acc: $81\%$\\
																													                     &                                       &  Long. CSF biomarkers +                        &                                       &                                    & Sens:$84\%$\\
																				                                       &                                       &  Long. Cognitive performance  +         &                                        &                                   & Spec:$80\%$ \\
\hline
Lin \textit{et al. }\cite{lin2018convolutional}&   2.5D CNN                &Baseline MRI +                         &    Regional          & MCI-to-AD &   AUC: $0.86$ \\
                                                                                 &    Multi-modal          & 325 Free Surfer feature        &       Hippocampus   & Conversion & Acc: $79.9\%$ \\
                                                                                 &    PCA + Lasso + NN          &         &          &  &                        Sens:$84.0\%$\\
																										                             &                                               &         &          &  &                        Spec:$74.8\%$\\
\hline
Lu \textit{et al. }\cite{lu2018multimodal}     &   Multi-modal              & MRI + FDG-PET                          &    Whole Brain     &  Stable MCI vs       & Acc: $82.9\%$\\
                                                                                 &   Multi-scale NN         &  Long. time-points                     &                               &  Progressive MCI  & Sens:79.7\% \\
																			                                           &           &              &                               &                                                                                                        & Spec:83.8\% \\

\hline
Esmaeilzadeh \textit{et al. }\cite{esmaeilzadeh2018end}     &   Multi-modal              & MRI +                            &   Whole Brain                  &  AD-NC-MCI      & Acc: $94.1\%$    \\
                                                                                                             &                                        &  Age + Gender             &                                            &  Classification  &  Sens:$94\%$\\
																																																	           &                                        &                                         &                                            &                            &  Sens:$91\%$\\
\hline
Moradi \textit{et al.} \cite{moradi2015machine}     &       Random Forest         & Baseline MRI + Age                                                &   Whole Brain                  &  MCI-to-AD     &  AUC: $0.9$  \\
                                                                                                &     Multi-modal             &  Baseline Cognitive measurements +                     &                                            &  Conversion  &   Acc: $82\%$ \\
																																													      &                                        & (RAVLT + ADAS-cog)                                                     &                                            &    & Sens: 87\% \\
																																						                    &                                        & (MMSE + CRD-SB + FAQ)                                              &                                            &    &  Spec:74\%\\
\hline 
Spasov \textit{et al.} \cite{spasov2019parameter}     &    3D-CNN                      & MRI +                                           &   Whole Brain                  &  MCI-to-AD     &  AUC: $0.925$  \\
                                                                                                &     Multi-modal             &  Demographic data +                     &                                            &  Conversion  &   Acc: $86\%$ \\
																																													      &                                        & neuropsychological data +           &                                            &    & Sens: 87.5\% \\
																																						                    &                                        & APOe4 genetic data                       &                                            &    &  Spec:85\%\\
 \end{tabular}
 \end{center}
 \caption{ML: Machine Learning, 2D: 2-Dimension, 3D: 3 Dimension, NN: Neural Network, CNN: Convolutional Neural Network, R-CNN: Recurrent Convolutional Neural Network, Acc: Accuracy, Sens: Sensitivity, Spec: Specificity, AUC: Area Under Curve. CSF: Cerebrospinal fluid, Long: Longitudinal, RAVLT: Rey's Auditory Verbal Learning Test, ADAS-cog: Alzheimer's Disease Assessment Scale cognitive subtest, MMSE: MiniMental State Examination, CDR-SB: Clinical Dementia Rating Sum of Boxes, FAQ: Functional Activities Questionnaire.}
 \label{Table:Literature}
 \end{table*}

\textit{This study.} Unlike prior studies, we investigate the feasibility of predicting the “rate of cognitive decline” in MCI patients at the first visit by processing only the baseline MRI and routinely collected clinical data. The training data is separated into two classes based on MMSE-rate values. We train our model with “slowly deteriorating/stable” or “rapidly deteriorating” classes formed based on MMSE-rate values. Therefore, we do not predict patients that convert to AD. However, some MCI cases deteriorated faster than the others. We investigate the prediction performance of our multi-modality architecture to predict the rapidly deteriorating cases based on information available at only the baseline visit. The proposed hybrid architecture jointly learns brain patterns and morphology from MRI sequences and additional information from the demographic data. To our knowledge, this is the first research study that investigates the feasibility of predicting the rate of cognitive decline by processing routine data collected at the first visit. 

We follow the same concatenation approach as in \cite{esmaeilzadeh2018end} which is proposed for disease detection. Another similar study proposed in \cite{lin2018convolutional} trained a convolutional neural network with local patches extracted from the hippocampus and combined the extracted information with FreeSurfer brain data. Our results are similar in that combining CNN features with scalar brain data features obtained with FreeSurfer increases the prediction performance. However, our study has differences, since our model (i) does not predict the MCI-to-AD conversion probability but instead predicts the rate of cognition deterioration in MCI patients by utilizing only the first-visit data; (ii) identifies patterns within the whole brain MRI instead of only the hippocampus, and (iii) uses limited FreeSurfer brain data (6 additional volume elements) compared with the brain data used in~\cite{lin2018convolutional} (325 additional data). 

Although, we roughly compare our study with MC-to-AD conversion studies, note that there are differences on approaching the problem. To our knowledge, this is the first study that predicts “slowly deteriorating/stable” or “rapidly deteriorating” classes by processing routinely collected baseline clinical and demographic data (Baseline MRI, Baseline MMSE, Scalar Volumetric data, Age, Gender, Education, Ethnicity, Race). The training data is built based on MMSE-rate values.  Therefore, how our study approach predicting progressive MCI is different than the previous studies. Also note that our method uses only baseline data, not the data that is not routinely asking during visits (e.g., APOe4 genetic data) or longitudinal data. 


\section{Conclusions and Discussion}~\label{Conclusion} In this study, we investigate whether a machine learning-based system can predict cognitive decline in MCI patients at the initial visit by processing routinely collected clinical data. Unlike other studies that focus on predicting MCI-to-AD conversion or AD/CN/MCI classification, we approach the problem as an \textit{early prediction of cognitive decline rate in MCI patients}. The ability to identify an individual’s cognitive decline rate potentially helps the clinician to develop early preventive treatment strategies.

We observed the performances of 3 models for the prediction of cognitive-decline class. Our results confirm that there is a correlation between the cognitive decline and the clinical data obtained at the first visit. The imaging model achieved $0.67$ AUC. By adding brain volume and demographic information as scalar values to the system, the performance increased to $0.70$ AUC. Processing brain volumes (from FreeSurfer brain data) and demographic information as scalar values provide similar results as the hybrid module performance. Even though patient's cognitive condition is mostly decided based on non-imaging clinical data (e.g., MMSE score, patient age) at the clinical visit, and MRI scans are generally collected to exclude other brain pathology, our results show that the structural MRI provides useful information related to the patient`s cognitive condition and may further contribute to the clinical evaluation and follow-up of patients with MCI. We have conducted experiments on Alzheimer’s Disease Neuroimaging Initiative (ADNI) dataset (c.f., Section~\ref{Data}) due to the availability of longitudinal MMSE scores and baseline clinical data. To our knowledge, there is not any available dataset that has a rich source of information regarding the Alzheimer’s Disease and its progression. However, the system needs to be further investigated and validated on an independent dataset. 

Our system performance is lower compared to the published studies that investigate MCI-to-AD conversion or AD/CN classification by modeling disease progression by processing longitudinal data obtained at several visits or by processing additional data that is not routinely obtained during visits (e.g., APOe4 genetic data). Note that predicting cognitive decline is more challenging than AD/CN classification due to the subtle nature of pathological changes~\cite{lin2018convolutional}. Moreover, our system processed only data that is routinely collected at the first visit, and thus makes predictions based on much less information compared to studies that incorporate follow-up data through time-sequence analysis. 

The clinical and demographic information contains additional data that contributes to the algorithm decision. In this study, we utilized CNN-based architecture as a classifier and combined the imaging features with non-imaging data at a dense layer in a straightforward way. To our knowledge, there is not any comprehensive study that investigates the best merging methods of different sources of information in CNN-based architecture, and it is an open research area.


\section*{Acknowledgments}
This research is supported by the Department of Radiology of The Ohio State University College of Medicine. In addition, the project is partially supported by a donation from the Edward J. DeBartolo, Jr. Family (Funding), Master Research Agreement with Siemens Healthineers (Technical Support), and Master Research Agreement with NVIDIA Corporation (Technical Support). 

Data collection and sharing for this project was funded by the Alzheimer's Disease Neuroimaging Initiative (ADNI) (National Institutes of Health Grant U01 AG024904) and DOD ADNI (Department of Defense award number W81XWH-12-2-0012). ADNI is funded by the National Institute on Aging, the National Institute of Biomedical Imaging and Bioengineering, and through generous contributions from the following: AbbVie, Alzheimer’s Association; Alzheimer’s Drug Discovery Foundation; Araclon Biotech; BioClinica, Inc.; Biogen; Bristol-Myers Squibb Company; CereSpir, Inc.; Cogstate; Eisai Inc.; Elan Pharmaceuticals, Inc.; Eli Lilly and Company; EuroImmun; F. Hoffmann-La Roche Ltd and its affiliated company Genentech, Inc.; Fujirebio; GE Healthcare; IXICO Ltd.; Janssen Alzheimer Immunotherapy Research and Development, LLC.; Johnson and Johnson Pharmaceutical Research and Development LLC.; Lumosity; Lundbeck; Merck and Co., Inc.; Meso Scale Diagnostics, LLC.; NeuroRx Research; Neurotrack Technologies; Novartis Pharmaceuticals Corporation; Pfizer Inc.; Piramal Imaging; Servier; Takeda Pharmaceutical Company; and Transition Therapeutics. The Canadian Institutes of Health Research is providing funds to support ADNI clinical sites in Canada. Private sector contributions are facilitated by the Foundation for the National Institutes of Health (www.fnih.org). The grantee organization is the Northern California Institute for Research and Education, and the study is coordinated by the Alzheimer’s Therapeutic Research Institute at the University of Southern California. ADNI data are disseminated by the Laboratory for Neuro Imaging at the University of Southern California.

Conflict of interest: The authors declare that they have no conflict of interest.

Human Participants: This article does not contain any studies with human participants or animals performed by any of the authors. This article does not contain patient data.

\bibliographystyle{elsarticle-num}
\bibliography{bib}

\begin{thebibliography}{10}
\expandafter\ifx\csname url\endcsname\relax
  \def\url#1{\texttt{#1}}\fi
\expandafter\ifx\csname urlprefix\endcsname\relax\def\urlprefix{URL }\fi
\expandafter\ifx\csname href\endcsname\relax
  \def\href#1#2{#2} \def\path#1{#1}\fi

\bibitem{markesbery2010neuropathologic}
W.~R. Markesbery, Neuropathologic alterations in mild cognitive impairment: a
  review, Journal of {A}lzheimer's Disease 19~(1) (2010) 221--228.

\bibitem{petersen1999mild}
R.~C. Petersen, G.~E. Smith, S.~C. Waring, R.~J. Ivnik, E.~G. Tangalos,
  E.~Kokmen, Mild cognitive impairment: clinical characterization and outcome,
  Archives of neurology 56~(3) (1999) 303--308.

\bibitem{qarni2019multifactor}
T.~Qarni, A.~Salardini, A multifactor approach to mild cognitive impairment,
  in: Seminars in neurology, Vol.~39, Thieme Medical Publishers, 2019, pp.
  179--187.

\bibitem{lee2019predicting}
G.~Lee, K.~Nho, B.~Kang, K.-A. Sohn, D.~Kim, Predicting {A}lzheimer’s disease
  progression using multi-modal deep learning approach, Scientific Reports
  9~(1) (2019) 1952.

\bibitem{cummings2007disease}
J.~L. Cummings, R.~Doody, C.~Clark, Disease-modifying therapies for
  {A}lzheimer's disease: challenges to early intervention, Neurology 69~(16)
  (2007) 1622--1634.

\bibitem{suk2013deep}
H.-I. Suk, D.~Shen, Deep learning-based feature representation for
  {A}{D}/{M}{C}{I} classification, in: International Conference on Medical
  Image Computing and Computer-Assisted Intervention, Springer, 2013, pp.
  583--590.

\bibitem{eskildsen2015structural}
S.~F. Eskildsen, P.~Coup{\'e}, V.~S. Fonov, J.~C. Pruessner, D.~L. Collins,
  {A}lzheimer's Disease Neuroimaging~Initiative, et~al., Structural imaging
  biomarkers of {A}lzheimer's disease: predicting disease progression,
  Neurobiology of aging 36 (2015) S23--S31.

\bibitem{moradi2015machine}
E.~Moradi, A.~Pepe, C.~Gaser, H.~Huttunen, J.~Tohka, {A}lzheimer's Disease
  Neuroimaging~Initiative, et~al., Machine learning framework for early
  {M}{R}{I}-based {A}lzheimer's conversion prediction in {M}{C}{I} subjects,
  Neuroimage 104 (2015) 398--412.

\bibitem{tabuas2016prognosis}
M.~T{\'a}buas-Pereira, I.~Baldeiras, D.~Duro, B.~Santiago, M.~Ribeiro,
  M.~Leit{\~a}o, C.~Oliveira, I.~Santana, Prognosis of early-onset vs.
  late-onset mild cognitive impairment: Comparison of conversion rates and its
  predictors, Geriatrics 1~(2) (2016) 11.

\bibitem{FreeSurfer}
Freesurfer, \url{https://surfer.nmr.mgh.harvard.edu//}, [Online; accessed 30
  November 2019].

\bibitem{arevalo2015mini}
I.~Arevalo-Rodriguez, N.~Smailagic, M.~R. i~Figuls, A.~Ciapponi,
  E.~Sanchez-Perez, A.~Giannakou, O.~L. Pedraza, X.~B. Cosp, S.~Cullum,
  Mini-{M}ental {S}tate {E}xamination ({M}{M}{S}{E}) for the detection of
  {A}lzheimer's disease and other dementias in people with mild cognitive
  impairment ({M}{C}{I}), Cochrane Database of Systematic Reviews~(3) (2015).

\bibitem{harrell2000severe}
L.~E. Harrell, D.~Marson, A.~Chatterjee, J.~A. Parrish, The severe mini-mental
  state examination: a new neuropsychologic instrument for the bedside
  assessment of severely impaired patients with {A}lzheimer disease, Alzheimer
  Disease \& Associated Disorders 14~(3) (2000) 168--175.

\bibitem{ADNI}
{A}lzheimer`s disease neuroimaging initiative,
  \url{http://adni.loni.usc.edu//}, [Online; accessed 24 November 2018].

\bibitem{esmaeilzadeh2018end}
S.~Esmaeilzadeh, D.~I. Belivanis, K.~M. Pohl, E.~Adeli, End-to-end
  {A}lzheimer’s disease diagnosis and biomarker identification, in:
  International Workshop on Machine Learning in Medical Imaging, Springer,
  2018, pp. 337--345.

\bibitem{ronneberger2015u}
O.~Ronneberger, P.~Fischer, T.~Brox, U-net: Convolutional networks for
  biomedical image segmentation, in: International Conference on Medical image
  computing and computer-assisted intervention, Springer, 2015, pp. 234--241.

\bibitem{Deepbrain}
Deepbrain - skull strip algorithm, \url{https://github.com/iitzco/deepbrain//},
  [Online; accessed 10 October 2019].

\bibitem{nyul2000new}
L.~G. Ny{\'u}l, J.~K. Udupa, X.~Zhang, New variants of a method of mri scale
  standardization, IEEE transactions on medical imaging 19~(2) (2000) 143--150.

\bibitem{ioffe2015batch}
S.~Ioffe, C.~Szegedy, Batch normalization: Accelerating deep network training
  by reducing internal covariate shift, arXiv preprint arXiv:1502.03167 (2015).

\bibitem{lecun2015deep}
Y.~LeCun, Y.~Bengio, G.~Hinton, Deep learning, Nature 521~(7553) (2015) 436.

\bibitem{srivastava2014dropout}
N.~Srivastava, G.~Hinton, A.~Krizhevsky, I.~Sutskever, R.~Salakhutdinov,
  Dropout: a simple way to prevent neural networks from overfitting, The
  Journal of Machine Learning Research 15~(1) (2014) 1929--1958.

\bibitem{litjens2017survey}
G.~Litjens, T.~Kooi, B.~E. Bejnordi, A.~A.~A. Setio, F.~Ciompi, M.~Ghafoorian,
  J.~A. Van Der~Laak, B.~Van~Ginneken, C.~I. S{\'a}nchez, A survey on deep
  learning in medical image analysis, Medical image analysis 42 (2017) 60--88.

\bibitem{arbabshirani2017single}
M.~R. Arbabshirani, S.~Plis, J.~Sui, V.~D. Calhoun, Single subject prediction
  of brain disorders in neuroimaging: promises and pitfalls, Neuroimage 145
  (2017) 137--165.

\bibitem{krogh1992simple}
A.~Krogh, J.~A. Hertz, A simple weight decay can improve generalization, in:
  Advances in neural information processing systems, 1992, pp. 950--957.

\bibitem{kingma2014adam}
D.~P. Kingma, J.~Ba, Adam: A method for stochastic optimization, arXiv preprint
  arXiv:1412.6980 (2014).

\bibitem{goodfellow2016deep}
I.~Goodfellow, Y.~Bengio, A.~Courville, Deep learning, MIT press, 2016.

\bibitem{van2006hippocampal}
L.~A. van~de Pol, A.~Hensel, W.~M. van~der Flier, P.~J. Visser, Y.~A.
  Pijnenburg, F.~Barkhof, H.~J. Gertz, P.~Scheltens, Hippocampal atrophy on
  {M}{R}{I} in frontotemporal lobar degeneration and {A}lzheimer’s disease,
  Journal of Neurology, Neurosurgery \& Psychiatry 77~(4) (2006) 439--442.

\bibitem{karas2004global}
G.~Karas, P.~Scheltens, S.~Rombouts, P.~Visser, R.~Van~Schijndel, N.~Fox,
  F.~Barkhof, Global and local gray matter loss in mild cognitive impairment
  and {A}lzheimer's disease, Neuroimage 23~(2) (2004) 708--716.

\bibitem{lin2018convolutional}
W.~Lin, T.~Tong, Q.~Gao, D.~Guo, X.~Du, Y.~Yang, G.~Guo, M.~Xiao, M.~Du, X.~Qu,
  et~al., Convolutional neural networks-based {M}{R}{I} image analysis for the
  {A}lzheimer’s disease prediction from mild cognitive impairment, Frontiers
  in neuroscience 12 (2018).

\bibitem{dimitriadis2018random}
S.~I. Dimitriadis, D.~Liparas, M.~N. Tsolaki, A.~D.~N. Initiative, et~al.,
  Random forest feature selection, fusion and ensemble strategy: Combining
  multiple morphological mri measures to discriminate among healhy elderly,
  mci, cmci and alzheimer’s disease patients: From the alzheimer’s disease
  neuroimaging initiative (adni) database, Journal of neuroscience methods 302
  (2018) 14--23.

\bibitem{long2017prediction}
X.~Long, L.~Chen, C.~Jiang, L.~Zhang, A.~D.~N. Initiative, et~al., Prediction
  and classification of alzheimer disease based on quantification of mri
  deformation, PloS one 12~(3) (2017).

\bibitem{vieira2017using}
S.~Vieira, W.~H. Pinaya, A.~Mechelli, Using deep learning to investigate the
  neuroimaging correlates of psychiatric and neurological disorders: Methods
  and applications, Neuroscience \& Biobehavioral Reviews 74 (2017) 58--75.

\bibitem{frisoni2010clinical}
G.~B. Frisoni, N.~C. Fox, C.~R. Jack, P.~Scheltens, P.~M. Thompson, The
  clinical use of structural mri in alzheimer disease, Nature Reviews Neurology
  6~(2) (2010) 67--77.

\bibitem{mcevoy2011mild}
L.~K. McEvoy, D.~Holland, D.~J. Hagler~Jr, C.~Fennema-Notestine, J.~B. Brewer,
  A.~M. Dale, Mild cognitive impairment: baseline and longitudinal structural
  mr imaging measures improve predictive prognosis, Radiology 259~(3) (2011)
  834--843.

\bibitem{rathore2017review}
S.~Rathore, M.~Habes, M.~A. Iftikhar, A.~Shacklett, C.~Davatzikos, A review on
  neuroimaging-based classification studies and associated feature extraction
  methods for alzheimer's disease and its prodromal stages, NeuroImage 155
  (2017) 530--548.

\bibitem{spasov2019parameter}
S.~Spasov, L.~Passamonti, A.~Duggento, P.~Li{\`o}, N.~Toschi, A.~D.~N.
  Initiative, et~al., A parameter-efficient deep learning approach to predict
  conversion from mild cognitive impairment to alzheimer's disease, Neuroimage
  189 (2019) 276--287.

\bibitem{qiu2018fusion}
S.~Qiu, G.~H. Chang, M.~Panagia, D.~M. Gopal, R.~Au, V.~B. Kolachalama, Fusion
  of deep learning models of mri scans, mini--mental state examination, and
  logical memory test enhances diagnosis of mild cognitive impairment,
  Alzheimer's \& Dementia: Diagnosis, Assessment \& Disease Monitoring 10
  (2018) 737--749.

\bibitem{lu2018multimodal}
D.~Lu, K.~Popuri, G.~W. Ding, R.~Balachandar, M.~F. Beg, Multimodal and
  multiscale deep neural networks for the early diagnosis of alzheimer’s
  disease using structural mr and fdg-pet images, Scientific reports 8~(1)
  (2018) 1--13.

\end{thebibliography}

\end{document}